\documentstyle[sprocl,psfig]{article}

\newcommand{\case}[2]{\mbox{\small $\displaystyle \frac{#1}{#2}$}}

\newcommand{\Pslash}{\mbox{$\not \! P$}}

\begin{document}

\title{Mesonic correlations and quark deconfinement}

\author{D. Blaschke}

\address{Fachbereich Physik, Universit\"at Rostock, D-18051 Rostock, Germany
\\E-mail: blaschke@darss.mpg.uni-rostock.de}

\author{P.C. Tandy}

\address{Center for Nuclear Research, Department of Physics, 
Kent State University\\ Kent, OH 44242, USA\\
E-mail: tandy@cnr2.kent.edu}  


\maketitle

\abstracts{A simple confining separable interaction Ansatz for the 
rainbow-ladder truncated QCD Dyson-Schwinger equations is used to implement 
chiral restoration and quark deconfinement in a study of
$\bar q q$ meson states at finite temperature.  The model is fixed at 
\mbox{$T=0$} by reproducing selected $\pi$ and $\rho$ properties. 
Deconfinement and chiral restoration are found to both occur at $T_c=146$ MeV.
In the pion sector, we investigate $M_\pi(T)$ and $f_\pi(T)$ along with 
the exact QCD mass relation and the GMOR relation.  
For the vector mode, we investigate the 3-space transverse 
and longitudinal masses $M_\rho^T(T)$ and $M_\rho^L(T)$, along with the width
for the decay \mbox{$\rho^0\rightarrow e^+e^-$}.
The equation of state (EOS) for the model is investigated in the 
\mbox{$T-\mu$} plane, a tri-critical point is identified and the relationship
to a bag model is discussed.   
The deconfinement transition in rapidly rotating neutron 
stars is considered and a new signal from the pulsar timing in binary 
systems with mass accretion is suggested.   The model is used to discuss the
possibility of a superconducting quark matter phase.}

\section{Introduction}
The experimental search for the QCD deconfinement phase transition in 
ultrarelativistic heavy-ion collisions will enter a new stage when the 
relativistic heavy-ion collider (RHIC) at Brookhaven will provide data 
complementary to those from the CERN SPS \cite{qm97}.
It is desirable to have a continuum field-theoretical modeling of quark 
deconfinement and chiral restoration at finite temperature and density (or
chemical potential $\mu$) that can be extended also to hadronic 
observables in a rapid and transparent way.  Significant steps in this
direction have recently been taken through a continuum approach to 
QCD$_{T,\mu}$ based on the truncated Dyson-Schwinger equations (DSEs) within 
the Matsubara formalism~\cite{bbkr,brs,mrs} and a recent review~\cite{br98}
is available.   A most appealing 
feature of this approach to modeling nonperturbative QCD$_{T,\mu}$ is 
that dynamical chiral symmetry breaking {\it and} confinement
is embodied in the the model gluon 2-point function constrained by chiral
observables at \mbox{$T=\mu=0$} and no new parameters are needed for
extension to \mbox{$T,\mu >0$}.   Approximations introduced by a 
specific truncation scheme for the set of DSEs can be systematically relaxed.
However due to the breaking of  $O(4)$ symmetry and the number of discrete 
Matsubara modes needed, the finite $T,\mu$ extension of realistic DSE models 
entails solution of a complicated set of coupled integral equations.   The 
generation of hadronic observables from such solutions, although a straightforward
adaption of the approach~\cite{T97rev} found to be successful at 
\mbox{$T=\mu=0$}, adds further
to the difficulties.  In the separable model we study here, detailed
realism is sacrificed in the hope that the dominant and essential features
may be captured in a simple and transparent format.   To this end we simplifiy an
existing \mbox{$T=\mu=0$} confining separable interaction Ansatz~\cite{b+} 
to produce a gaussian separable model for \mbox{$T,\mu >0$}~\cite{bkt}.  

\section{Confining separable Dyson-Schwinger equation model}

In a Feynman-like gauge where we take $D_{\mu\nu}=\delta_{\mu\nu}D(p-q)$
to be the effective interaction between quark colored vector currents, 
the rainbow approximation to the DSE for the quark propagator 
$S(p)=[i \rlap/p A(p) + B(p) + m_0]^{-1}$ yields in Euclidean metric
\begin{eqnarray}
B(p) &=& 
\frac{16}{3} \int \frac{d^4q}{(2\pi)^4} D(p-q) 
\frac{B(q)+m_0}{q^2A^2(q)+\left[ B(q)+m_0\right]^2} \,\,\, , \\
\left[ A(p)-1 \right]  p^2 &=& 
\frac{8}{3} \int \frac{d^4q}{(2\pi)^4} D(p-q) 
\frac{(p\cdot q) A(q)}{q^2A^2(q)+\left[ B(q)+m_0\right]^2} \,\,\, .
\end{eqnarray} 
We study a separable interaction given by~\cite{b+}
\begin{equation}
D(p-q) = D_0~ f_0(p^2)f_0(q^2) + D_1~ f_1(p^2)(p\cdot q)f_1(q^2)~,  
\label{model}
\end{equation}
where $D_0, D_1$ are strength parameters and the form factors, for 
simplicity, are here taken to be
$f_i(p^2) = \mbox{exp}(-p^2/\Lambda_i^2)$ with range parameters $\Lambda_i$.
It is easily verified that  if $D_0$ is non-zero, then 
\mbox{$B(p)=\Delta m~f_0(p^2)$}, and if $D_1$ is non-zero, then  
\mbox{$A(p)=1+\Delta a~f_1(p^2)$}.   The DSE then reduces to nonlinear
equations for the constants $\Delta m$ and $\Delta a$.
The form factors should be chosen to
simulate the $p^2$ dependence of $A(p)$ and $B(p)$ from a more realistic 
interaction.   We restrict our considerations here to the rank-1 case 
where $D_1=0$ and \mbox{$A(p)=1$}.  The parameters $D_0$, $\Lambda_0$ 
and $m_0$ are used to produce reasonable $\pi$ and $\omega$ properties
as well as to ensure the produced  $B(p)$ has a reasonable strength with
a range $\Lambda_0 \sim 0.6 \dots 0.8$ GeV  to be realistic~\cite{b+}.

If there are no solutions to $p^2A^2(p)+(B(p)+m_0)^2=0$
for real $p^2$ then the quarks are confined. If in the chiral limit 
($m_0=0$) there is a nontrivial solution for $B(p)$, then chiral symmetry 
is dynamical broken.  Both phenomena can be implemented in this separable model.
In the chiral limit, the model is confining if $D_0$ is strong enough to 
make $\Delta m/\Lambda_0\ge1/\sqrt{2{\rm e}}$.  Thus for a typical range
$\Lambda_0$,  confinement will typically occur with 
$M(p\approx 0)\ge 300$ MeV. 
 
Mesons as $q \bar q$ bound states are described by the Bethe-Salpeter 
equation which in the ladder  approximation for the present approach is
\begin{equation}
- \lambda(P^2) \Gamma(p,P) 
= \case{4}{3} \int \frac{d^4q}{(2\pi)^4} 
D(p-q) \gamma_\mu S(q_+) \Gamma(q,P)S(q_-) \gamma_\mu~,
\label{bs}
\end{equation}
where \mbox{$q_\pm = q \pm P/2$} and $P$ is the meson momentum.
The meson mass is identified from \mbox{$\lambda(P^2=-M^2)=$}$1$.
With the rank-1 separable interaction, only the $\gamma_5$ and the 
$\gamma_5 \Pslash$
covariants contribute
to the $\pi$~\cite{b+},  and here we retain only the dominant term
$\Gamma_\pi(p,P) = i \gamma_5 E_\pi (p,P)$.  For the vector meson, the 
only surviving form is \mbox{$\Gamma_{\rho \mu}(p,P) =$}
\mbox{$\gamma_\mu^T(P) E_\rho (p,P)$}, with $ \gamma_\mu^T(P)$ being the
projection of 
$\gamma_\mu$ transverse to $P$.   The separable solutions have the form 
$E_i (p,P)=f_0(p^2) C_i(P^2), ~~i=\pi, \rho~$, where the $C_i$ factor out
from Eq.~(\ref{bs}).  

In the limit where a zero momentum range for the interaction is simulated 
by \mbox{$f_0^2(q^2)\propto $}\mbox{$ \delta^4(q)$}, then the
expressions for the various BSE eigenvalues $\lambda(P^2)$ reduce to
those of the Munczek and Nemirovsky~\cite{mn} model which implements
extreme infrared dominance via  \mbox{$D(p-q) \propto \delta^{(4)}(p-q)$}. 
The correspondence is not complete because the quark DSE solution in this 
model has $A(p)\neq 1$.  The \mbox{$T,\mu >0$} generalization of this infrared
dominant (ID) model have been studied recently~\cite{brs,mrs}.
 
\section{Pion and rho-meson properties}
\label{sec:mesons}

With parameters $m_0/\Lambda_0=0.0096$, $D_0 \Lambda_0^2=128$ and 
$\Lambda_0=0.687$ GeV, the present Gaussian separable (GSM) model yields  
$M_\pi=0.14$ GeV, $M_\rho=M_\omega=0.783$ GeV, $f_\pi=0.104$ GeV, a chiral 
quark condensate $\langle \bar q q\rangle^{1/3}=-0.248$ GeV, and a 
$\rho-\gamma$ coupling constant $g_\rho=5.04$.

The generalization to $T\neq 0$ is systematically accomplished by 
transcription of the Euclidean quark 4-momentum via \mbox{$q \rightarrow$}
\mbox{$ q_n =$} \mbox{$(\omega_n, \vec{q})$}, where 
\mbox{$\omega_n=(2n+1)\pi T$} are the discrete Matsubara frequencies. 
The obtained $T$-dependence of the mass gap $\Delta m(T)$ allows for a study
of the deconfinement and chiral restoration features of this model. 
We find that both occur at \mbox{$T_c=$} 146~MeV where, in the chiral limit,
both $\Delta m(T)$ and $\langle \bar{q}q \rangle^0$ vanish sharply as 
\mbox{$(1-T/T_c)^\beta$} with the critical exponent having the mean
field value \mbox{$\beta=1/2$}. 

For the $\bar q q$ meson modes, the $O(4)$ symmetry is broken and the type of
mass shell condition employed must be specified.  If there is a bound  state, 
the associated pole contribution to the relevant $\bar q q$ propagator or 
current correlator will have a denominator proportional to 
\begin{equation}
1-\lambda(\Omega_m^2, \vec{P}^2) \; \propto  \; 
\Omega_m^2 + \vec{P}^2 +M^2(T)~.
\label{dennom}
\end{equation}
We investigate the meson mode eigenvalues $\lambda$ using only the
lowest meson Matsubara mode ($\Omega_m=0$) and the continuation
\mbox{$\vec{P}^2\longrightarrow -M^2$}.   The masses so identified are
spatial screening masses corresponding to a behavior $\exp(-M x)$ in the
conjugate 3-space coordinate $x$ and should correspond to the lowest bound
state if one exists.   

\begin{figure}[h]
\centerline{
\psfig{figure=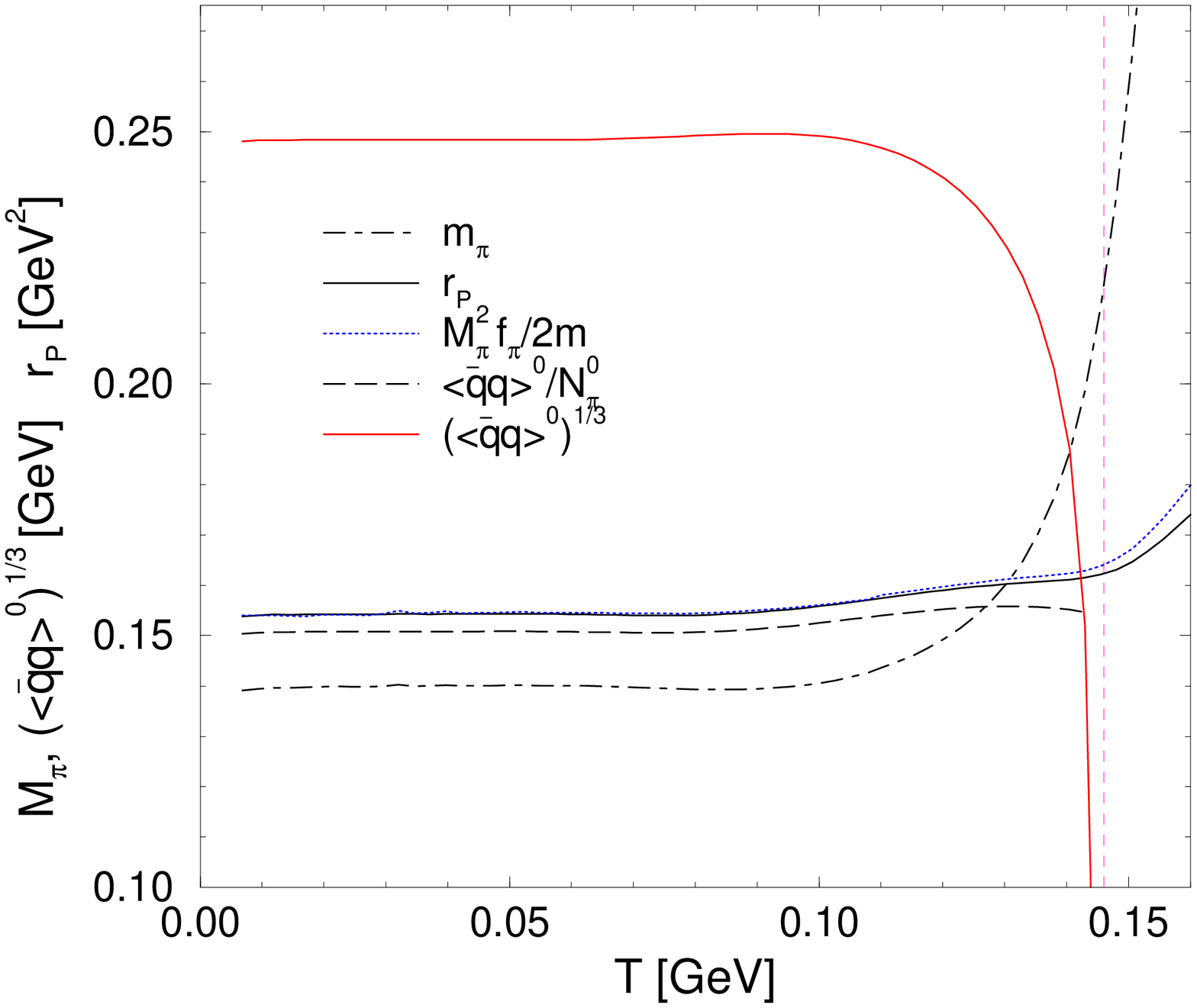,width=5cm,height=5.8cm,angle=-90}
\psfig{figure=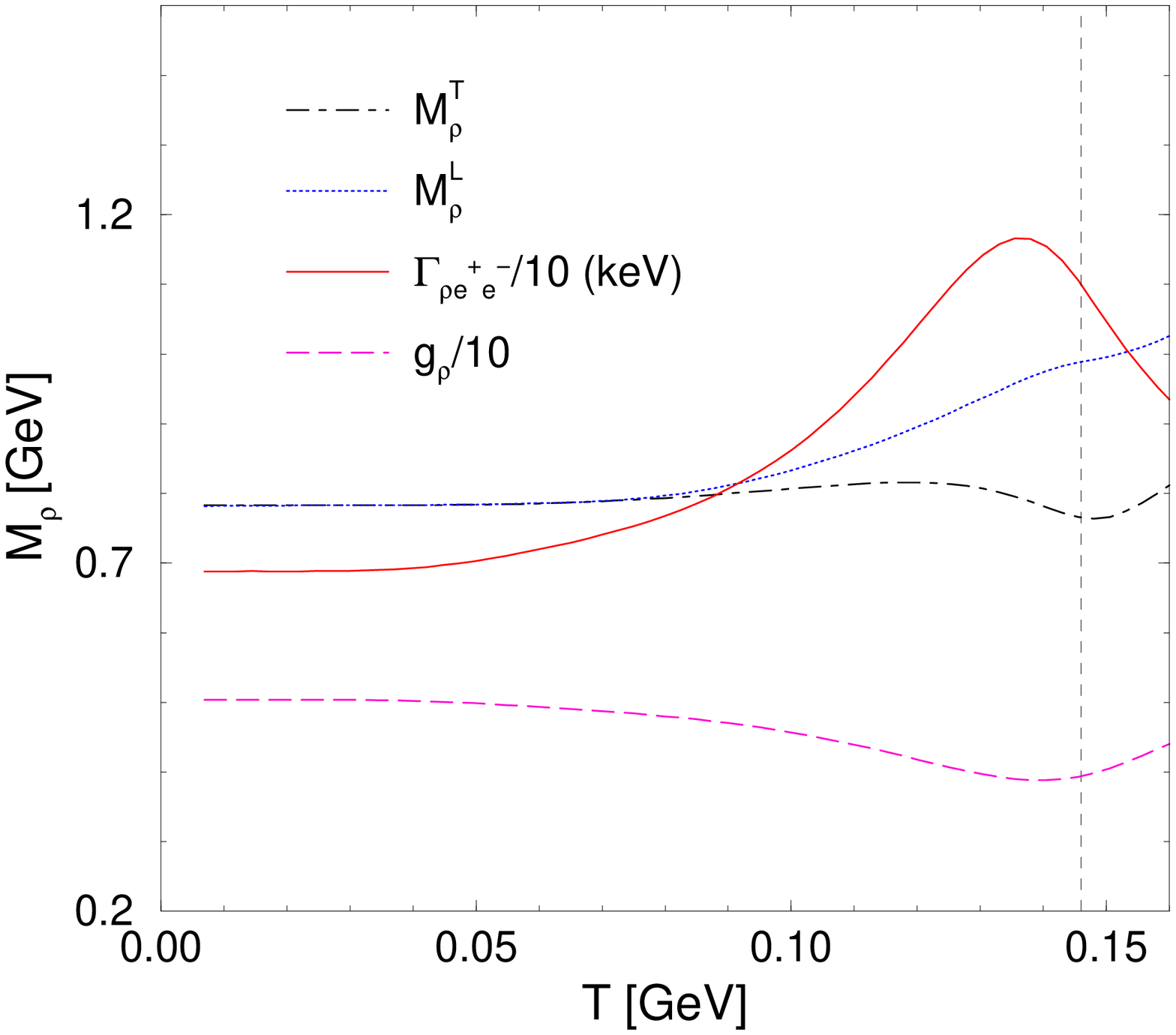,width=5cm,height=5.8cm,angle=-90}
}
\caption{\label{pirho} $T$-dependence of $\bar q q$ meson properties for 
the rank-1 separable model up to \mbox{$T_c=146$}~MeV.  
Left panel: Spatial $M_\pi(T)$ and quantities associated with
mass relations as dictated by chiral symmetry; Right panel: Spatial masses
$M_\rho^T(T)$ and $M_\rho^L(T)$ along with the \mbox{$\rho^0 \rightarrow e^+ e^-$} 
decay width and the associated vector coupling constant $g_{\rho}(T)$. }
\end{figure}
The obtained $\pi$ and $\rho$ masses displayed in Fig.~\ref{pirho} and are 
seen to be only weakly  $T$-dependent until near \mbox{$T_c=146$}~MeV.  This 
result for $M_\pi (T)$
reproduces the similar behavior obtained from the ladder-rainbow truncation
of the DSE-BSE complex with a more realistic interaction~\cite{bbkr}.  The 
qualitative behavior obtained for the 3-space 
transverse and longitudinal masses $M_\rho^T(T), M_\rho^L(T)$ agrees with 
that reported~\cite{mrs} for the limiting case of the zero momentum range
or ID model.  

To explore the extent to which the model respects the detailed constraints 
from chiral symmetry, we investigate the exact QCD  pseudoscalar mass 
relation~\cite{MRT98} which, after extension to \mbox{$T>0$}, is
\begin{equation}
\label{gen-GMOR}
M_\pi^2(T) \, f_\pi(T) = 2 m_0 \, r_P(T)~.
\end{equation}
Here $r_P$ is the residue at the pion pole in the pseudoscalar 
vertex, and in the present model, is given by
\begin{equation}
\label{rp}
i r_P(T) =  N_c \; T \sum_n {\rm tr}_s \int \frac{d^3q}{(2\pi)^3}\, 
     \gamma_5 S(q_n+\case{\vec{P}}{2}) \Gamma_\pi (q_n;\vec{P}) 
                                   S(q_n-\case{\vec{P}}{2})\,.
\end{equation}
The relation in Eq.~(\ref{gen-GMOR}) is a consequence of the pion pole structure
of the isovector axial Ward identity which links the quark propagator, the 
pseudoscalar vertex and the axial vector vertex~\cite{MRT98}.  In the chiral 
limit, \mbox{$r_P \rightarrow$} 
\mbox{$ \langle \bar{q} q\rangle^0/f_\pi^0$} and Eq.(\ref{gen-GMOR}), 
for small mass, produces the Gell-Mann--Oakes--Renner (GMOR) relation.  

The exact mass relation, Eq.~(\ref{gen-GMOR}), can only be approximately 
satisfied when the various quantities are obtained approximately such as 
in the present separable model.  The error can be used to assess the 
reliability of the present 
approach to modeling the behavior of the pseudoscalar bound state as the
temperature is raised towards $T_c$.    Our findings are displayed in 
Fig.~\ref{pirho}.  There the solid line represents $r_P(T)$ calculated from
the quark loop integral in Eq.~(\ref{rp}); the dotted line represents  $r_P$ 
extracted from  the other quantities in Eq.~(\ref{gen-GMOR}).  It is 
surprising that the separable model obeys this exact QCD mass relation to better
than 1\% for the complete temperature range up to the restoration of chiral
symmetry.    Also evident from Fig.~\ref{pirho} is that $M_\pi(T)$ and the chiral
condensate are largely temperature independent until within about $0.8~T_c$
whereafter $M_\pi$ rises roughly as fast as the condensate falls.  

We have also investigated the (approximate) GMOR relation for the present model.
The quantity \mbox{$ \langle \bar{q} q\rangle^0/N_\pi^0$} is displayed in
Fig.~\ref{pirho} as the long-dashed line, and if the GMOR relation were exactly
obeyed, this would coincide with \mbox{$M_\pi^2\,f_\pi/2m$} which is the dotted
line.  The quantity $N_\pi^0$ enters here via its role as the normalization 
constant of the chiral limit $\pi$ BS amplitude \mbox{$E_\pi^0(p^2)=$}
\mbox{$i\gamma_5 B_0(p^2)/N_\pi^0$}.  If all covariants for the pion were to be 
retained and the axial vector Ward identity were obeyed, one would have
\mbox{$N_\pi=f_\pi$}~\cite{MRT98}.   
The results in Fig.~\ref{pirho} indicate that the GMOR relation contains an 
error of about $5$\% when compared either to the exact mass relation or to
the quantities produced by the separable model and that this is 
temperature-independent until about $0.9~T_c$.   It should be 
noted that $f_\pi^0, N_\pi^0$ and $ \langle \bar{q} q\rangle^0$ are equivalent
order parameters near $T_c$ and have weak $T$-dependence below $T_c$.      
A consequence is that $M_\pi^2 \, f_\pi$, $r_P$ and 
\mbox{$ \langle \bar{q} q\rangle^0/N_\pi^0$} are almost $T$-independent and so 
are the estimated errors for the two mass relations linking these quantities.  
Since we  obtain $M_\pi$ and $f_\pi$ from the model BSE solutions at finite
current quark mass, $f_\pi$ does not exactly decrease to zero and 
$M_\pi$ does not exactly diverge at $T_c$. 

Vector mesons play an important role as precursors to di-lepton events in
relativistic heavy-ion collisions and it is important to explore the intrinsic
$T$-dependence of electromagnetic and leptonic vector coupling constants that
can arise from the quark-gluon dynamics that underlies the finite extent of
the vector $\bar q q$ modes.  The present model provides
a simple framework for such investigations. The electromagnetic decay constant 
$g_\rho(T)$ that describes the coupling
of the transverse $\rho^0$ to the photon is given by~\cite{IKR99}
\begin{eqnarray}
   \frac{{M^T_\rho}^2(T)}{g_\rho(T)} &=& \case{N_c}{3}\,
        T \sum_n {\rm tr}_s \int \frac{d^3q}{(2\pi)^3}\, 
        \gamma_\mu S(q_n+\case{\vec{P}}{2}) 
        \Gamma_\mu^T(q_n;\vec{P}) S(q_n-\case{\vec{P}}{2})~,
\label{rhophoton}
\end{eqnarray}
after accounting for the normalization~\cite{T97rev} of the present BS amplitudes.
The electromagnetic decay width of the transverse $\rho$ mode is calculated
from
\begin{eqnarray}
        \Gamma_{\rho^0 \rightarrow e^+\,e^-}(T) &=& 
                \frac{4\pi\,\alpha^2\,M^T_\rho(T)}{3\;g_\rho^2(T)}~.
\end{eqnarray}
At \mbox{$T=0$} the experimental value is
\mbox{$\Gamma_{\rho^0\rightarrow e^+e^-}(0) =$}
\mbox{$6.77$}~keV corresponding to the value \mbox{$g_\rho(0) =5.03$}.  Our
results for $g_\rho(T)$ and \mbox{$\Gamma_{\rho^0\rightarrow e^+e^-}(T)$}
are displayed in Fig.~\ref{pirho}.  This electromagnetic width of the 3-space
transverse $\rho$ increases with $T$ and reaches a maximum of 1.72 times the 
$T=0$ width at about $0.9~T_c$.    An increasing electromagnetic width for the 
$\rho$ has been found empirically to be one of the possible medium effects 
that influence the heavy-ion dilepton spectrum~\cite{sb}.

\section{Equation of state (EOS) for quark matter}
\begin{figure}[ht]
\centerline{
\psfig{figure=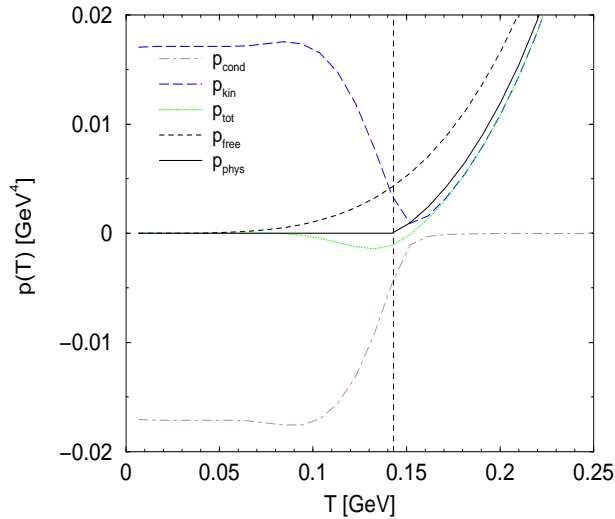,width=8cm,height=7cm,angle=-90}
}
\caption{\label{pres_T}The quark matter pressure as a function of temperature 
for the separable confining model. The kinetic part (long dashed line) is 
overcompensated by the condensate part (dot-dashed line) resulting in an
instability ${\rm d} (p_{\rm cond}+p_{\rm kin})/ {\rm d}  T < 0$ (dotted line) 
for $T<T_c=146$~MeV 
which is characteristic for confining quark models. The physical pressure
of quark matter (solid line) vanishes in this region. For comparison, the
free quark matter pressure is shown by the dashed line.}
\end{figure}

\begin{figure}[ht]
\centerline{
\psfig{figure=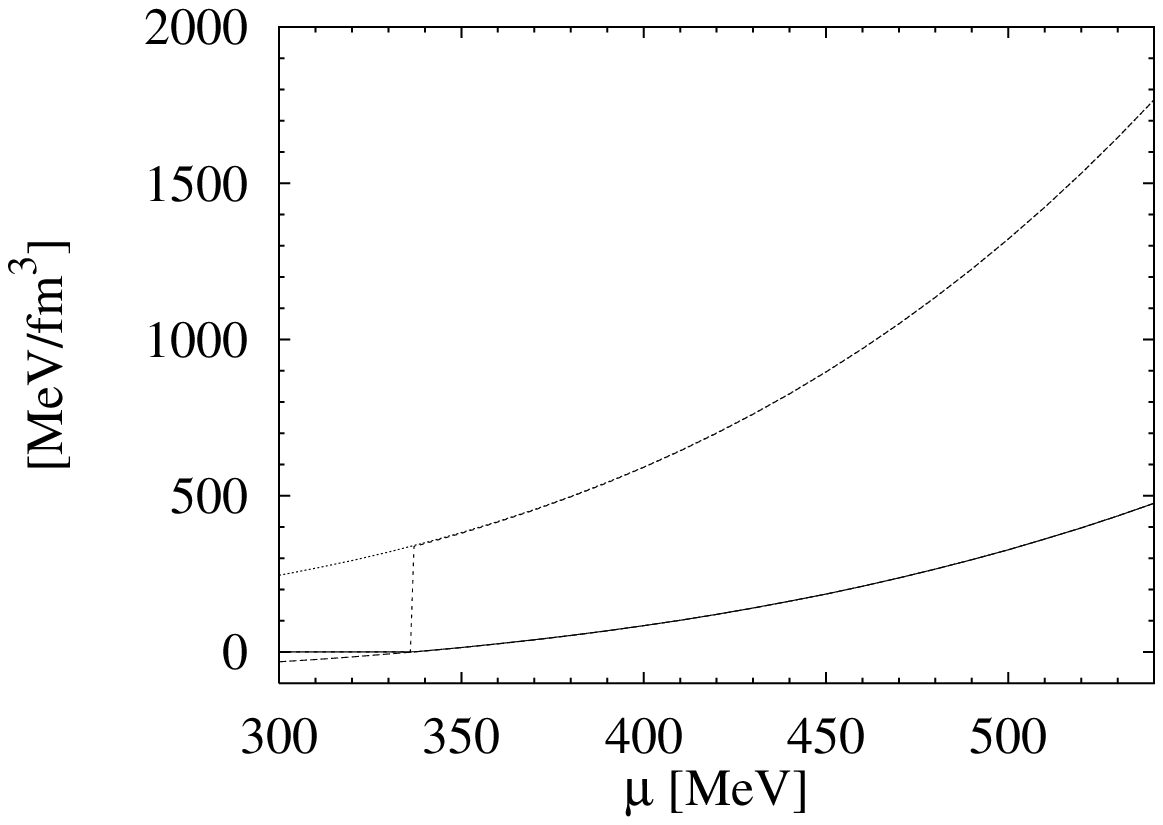,width=6cm,height=5cm,angle=0}
\psfig{figure=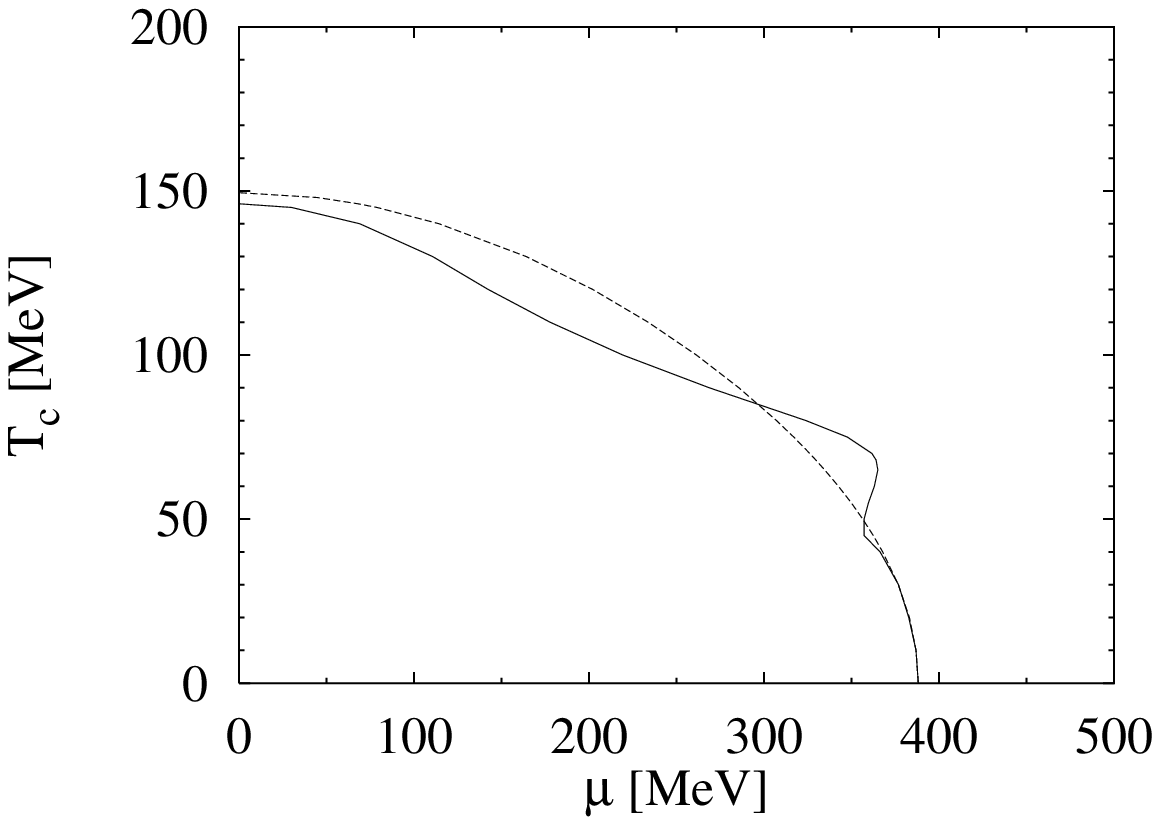,width=6cm,height=5cm,angle=0}
}
\caption{\label{eos}Left panel: pressure (solid line) and energy density 
(long dashed line) vs. chemical potential at \mbox{$T=0$} for the GSM with 
$\Lambda_0=0.756$ GeV and $D_0\Lambda_0^2=97.0$ in the chiral limit. 
The phase transition from the confining quark matter phase to the deconfined 
one with restored chiral symmetry occurs at $\mu=0.337$ GeV and is first order.
The results are coincident with a bag model EOS (dashed and dotted lines) for
a bag constant $B=75$ MeV/fm$^3$.
Right panel: quark matter phase diagram for the GSM with $\Lambda_0=0.687$ GeV 
and $D_0\Lambda_0^2=128$. Along the dashed line the pressure vanishes, the 
solid line separates the chiral symmetric phase from the broken one. 
A tricritical point is obtained at $T\sim 127$ MeV, $\mu\sim 120$ MeV.}
\end{figure}

The thermodynamical properties of the confining quark model and in particular
the EOS and the phase diagram can be obtained from the grand canonical 
thermodynamical potential $\Omega(T,V,\mu)=T \ln Z(T,V,\mu)=-p(T,\mu)V$, where 
the contributions to the pressure (for a homogeneous system) 
\begin{equation}
\label{ptot}
p(T,\mu)=p_{\rm cond}(T,\mu)+p_{\rm kin}(T,\mu)+p_0 
\end{equation}
are obtained from a mean-field approximation to the Euclidean path integral 
representation of the grand canonical partition function $Z(T,V,\mu)$.
In the rank-one separable gluon propagator model, the condensate contribution
is $p_{\rm cond}(T,\mu)=3\,\Delta m(T,\mu)^2/(16\,D_0)$ and the  kinetic part
of the quark pressure is given by
\begin{equation}
\label{pkin}
p_{\rm kin}(T,\mu)=2 N_c N_f \int\frac{d^3 k}{(2\pi)^3} T \sum_{n}
\ln \left(\frac{k_n^2+M^2(k_n^2)}{k_n^2+m_0^2}\right)+p_{\rm free}(T,\mu)~.
\end{equation}
In Eq. (\ref{pkin}), $k_n^2=[(2n+1)\pi T + i \mu]^2 + {\bf k}^2$, 
$M(k_n^2)=m_0+\Delta m(T,\mu) f_0(k_n^2)$ and  
the divergent 3-momentum integration has been regularized
by subtracting the free quark pressure and adding it in the well-known 
\cite{kapusta} regularized form $p_{\rm free}(T,\mu)$.
The pressure contribution $p_0$ is found such that the total pressure 
(\ref{ptot}) at the phase boundary in the $T,\mu$-plane vanishes, 
see Fig. \ref{eos}. 
While investigating this EOS for the separable confining quark model defined
above we have observed that, as a function of the coupling parameter $D_0$,
an instability ${\rm d}(p_{\rm cond}+p_{\rm kin})/{\rm d}T<0$ 
occurs when the criterion for confinement (absence of quasiparticle mass poles)
is fulfilled. The physical quark pressure in the confinement domain of the 
phase diagram vanishes, see  Fig. \ref{pres_T}. 
The results for the EOS and the phase diagram can be compared to those for the 
zero momentum range model \cite{brs} with the important modification that in 
the present finite range model the tricritical point is obtained at finite 
chemical potential whereas with zero range
it was found on the $\mu=0$ axis of the phase diagram, see Fig. \ref{eos}. 
The location of the tricitical point could be experimentally verified in 
CERN-SPS experiments provided that changes in the pion momentum correlation 
function could be detected as a function of the beam energy \cite{misha}.
A particularly interesting phenomenological application is
the $T=0$ quark matter EOS which is a necessary element for
studies of quark deconfinement in neutron stars.
The present Gaussian separable model 
leads~\cite{bb99} to a bag model EOS for quark matter with a bag constant  
$B(T=0)=150$ MeV/fm$^3$ for the parameter set ($D_0~\Lambda_0^2=128$) employed 
in Sec.~\ref{sec:mesons}.   A second parameter set ($D_0~\Lambda_0^2=97$) that is also 
confining and provides an equally good description of the same $\pi$ and $\rho$ 
properties produces $B(T=0)=75$ MeV/fm$^3$;  
see also Fig. \ref{eos}.  More stringent constraints on the low-temperature EOS 
will require the inclusion of hadronic excitations including the nucleon. 
\section{Deconfinement in rotating neutron stars}

For the discussion of deconfinement in neutron stars, it is crucial to go 
beyond the mean field description and to include into the EOS also the hadronic
bound states (neutrons, protons, mesons) in the confined phase. 
This task has not yet 
been solved and therefore, we adopt for this phase a Walecka model as it is 
introduced in \cite{kapusta}.  In constructing the phase 
transition to the deconfined quark matter as described by the GSM with a 
parameter set leading to $B=75$ MeV/fm$^3$ we have to obey the constraints of
global baryon number conservation and charge neutrality \cite{glendenning}. 
The composition of the neutron star matter is also constrained by the processes
which establish $\beta-$ equilibrium in the quark phase ($d\to u+e^-+\bar\nu$)
and in the hadronic phase ($n\to p+e^-+\bar\nu$). For the given EOS we 
obtain a deconfinement transition of first order where the hadronic phase
is separated from the quark matter one by a mixed phase in the density interval
$1.39\le n/n_0\le 2.37$, where $n_0=0.16~{\rm fm}^{-3}$ is the nuclear 
saturation density~\cite{c+}.

\begin{figure}[ht]
\centerline{
\psfig{figure=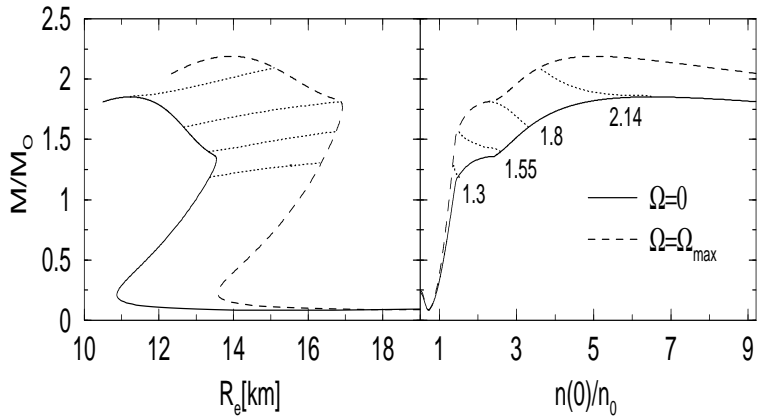,width=10cm,height=5.8cm,angle=0}
}
\caption{\label{starmass} The mass $M$ as a function of the equatorial radius 
$R_e$ (left panel)
and the central density (right panel) for a neutron star with deconfinement
transition. Rotating configurations (dashed lines) with maximum rotation 
frequency are connected with the static ones (solid lines) by lines of 
constant baryon number $N/N_\odot=1.3,~1.55,~1.8,~2.14$, respectively.
The occurence of an extended quark matter core in the compact star is 
compatible with recently derived constraints on maximum mass and radius from
QPO observations in low-mass X-ray binaries, see text.}
\end{figure}

\begin{figure}[ht]
\centerline{
\psfig{figure=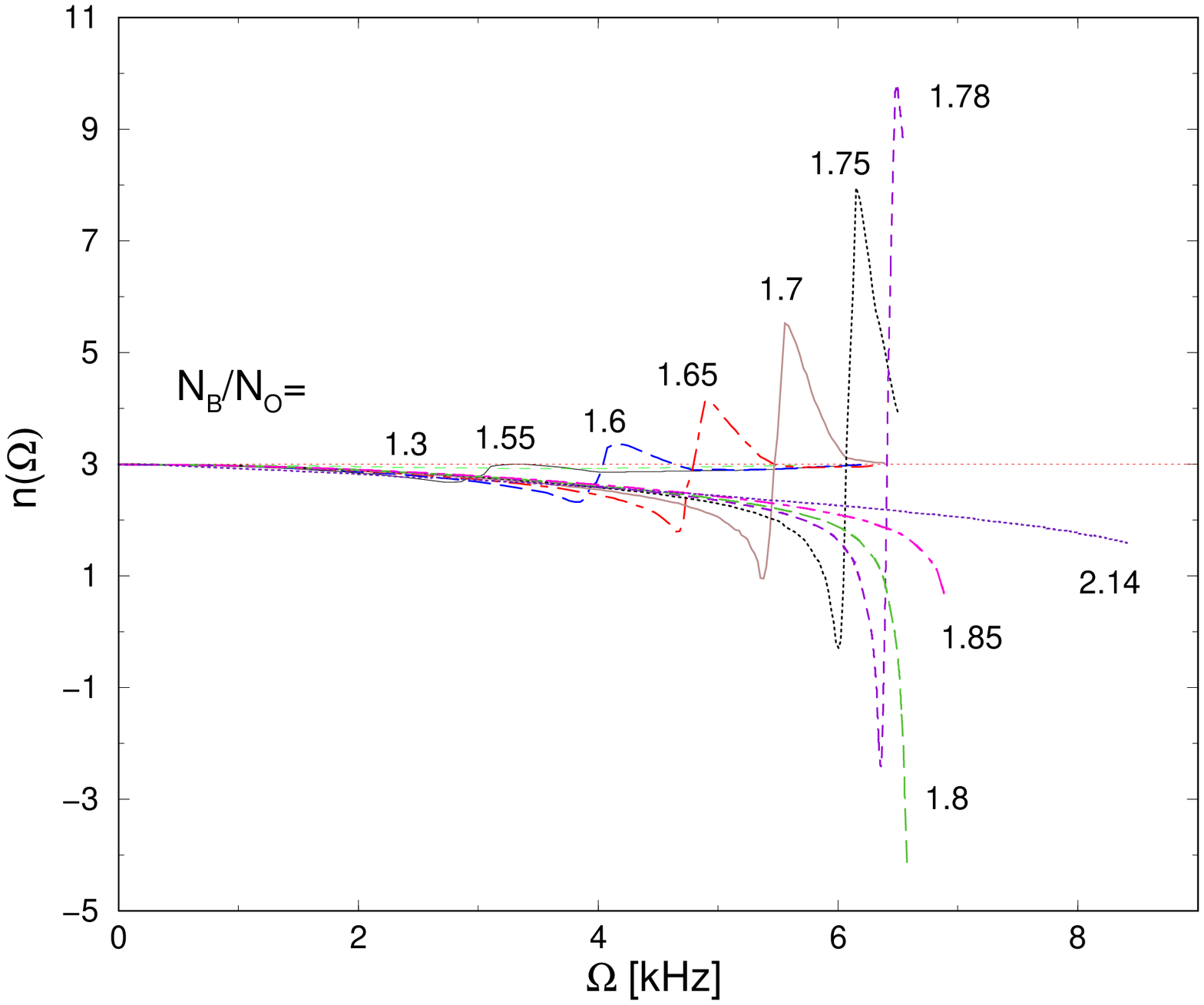,width=5cm,height=5.8cm,angle=-90}
\psfig{figure=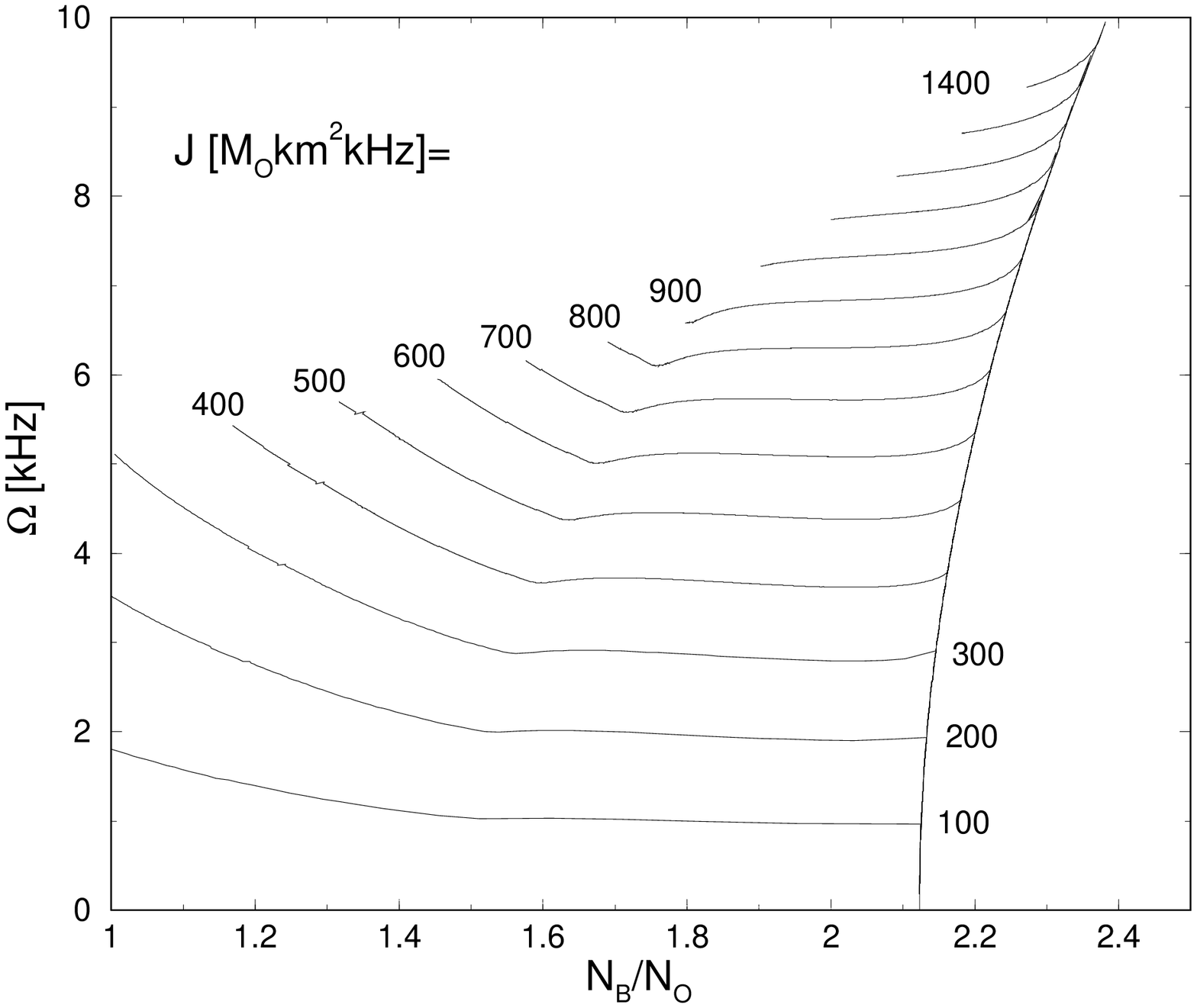,width=5cm,height=5.8cm,angle=-90}
}
\caption{\label{spin} Deconfinement signals from the pulse timing of rapidly 
rotating 
compact stars. Left panel: The braking index deviates from $n=3$ when during 
the spin-down evolution of a pulsar a quark matter core is formed which 
entails a change in the moment of inertia. The larger the radius of the quark 
core, the more pronounced the signal. 
Right panel: The spin-down evolution of a compact star with mass accretion 
at constant angular momentum $J$ flips to a spin-up behaviour at the onset of 
deconfinement.}
\end{figure}

All observable consequences should be discussed for fastly rotating compact 
objects and therefore we have studied these rotating configurations using this
model-EOS with a deconfinement transition. The result is shown in Fig. 
\ref{starmass} and shows that within the present approach a deconfinement 
transition in compact stars is compatible with constraints on radius and mass
recently derived from the observation of QPO in low mass X-ray binaries
(LMXBs)~\cite{lamb}. 
The basic quantity for the study of a deconfinement transition rotating 
compact stars is the moment of inertia which governs the rotation and thus 
the spin-down characteristics. Changes in the angular velocity $\Omega(t)$ 
as a function of time can occur, e.g., due to magnetic dipole 
radiation~\cite{frido} or mass accretion~\cite{c+}. 
During the time evolution of an isolated pulsar, the deviation of the braking
index from $n(\Omega)=3$ can signal not only the occurence of a quark matter 
core, but also its size~\cite{c+}. 
We have found that in LMXBs with mass accretion at conserved angular momentum  
the occurence of a quark matter core would reflect itself in a change from a 
spin-down to spin-up era~\cite{c+}, see Fig. \ref{spin}. 
More detailed investigations of these scenarios have to be performed with a
more realistic EOS, in particular for the hadronic phase. The possibility of 
a $T=0$ quark matter EOS which corresponds to a bag model EOS with small bag
constants of the order of $B=70$ MeV/fm$^{3}$ is an important result of the
study of the confining quark model which bears interesting consequences for
the study of further nontrivial phases in high-density QCD, as e.g. (color-)
superconductivity. 

\section{Superconducting quark matter}
The possible occurence of a superconducting quark matter phase~\cite{bl84}
has been recently reconsidered on the basis of nonperturbative approaches to
the effective quark 4-point interaction~\cite{br98,alford,rapp,carter} and 
critical temperatures of the order of $50$ MeV with diquark pairing gaps of
$\approx 100$ MeV have been obtained.
So, if quark matter occurs in the interior of compact stars as advocated in the
previous section, then it had to be realised in such a superconducting phase.
Deconfinement in compact stars can thus result in effects on the 
magnetic field structure~\cite{bss} as well as the cooling curves of pulsars.
Contrary to previous estimates~\cite{bl84}, low temperature quark matter is a 
superconductor of second kind and thus the magnetic field can penetrate into 
the quark core in Abrikosov vortices and does not decay at timescales shorter
than $10^7$ years. Thus the occurence of superconducting quark matter phase in
compact stars does not contradict observational data~\cite{bss}. 
The recently developed nonperturbative approaches to diquark condensates in 
high-density quark matter~\cite{alford,rapp,carter} can be further constrained 
by studying the consequences for cooling curves of pulsars~\cite{BKSV} which 
have to be consistent with the observational data~\cite{tsuruta}.

\section{Conclusions}
A simple confining separable interaction Ansatz for the rainbow-ladder 
truncated QCD Dyson-Schwinger equations is found capable of modeling 
$\bar q q$ meson states at \mbox{$T>0$} together with quark deconfinement and chiral 
restoration.  Deconfinement and chiral restoration are found to both occur at 
$T_c=146$ MeV.  The spatial screening masses for the meson modes are obtained.
We find that, until near $T_c$, $M_\pi(T)$ and $f_\pi(T)$ are weakly 
$T$-dependent and that this model obeys the exact QCD pseudoscalar mass relation 
to better than 1\%.   The GMOR relation is found to be accurate to within 5\% until
very near $T_c$.  For the vector mode, the 3-space transverse 
and longitudinal masses $M_\rho^T(T)$ and $M_\rho^L(T)$ are weakly $T$-dependent
while the width for the electromagnetic decay \mbox{$\rho^0\rightarrow e^+e^-$} is 
found to increase to 1.72 times the \mbox{$T=0$} width.
The equation of state (EOS) for the model is investigated in the $T-\mu$ plane
and it shows a tricritical point at $T= 127~{\rm MeV}, ~\mu= 120~{\rm MeV}$.
At $T=0$ the EOS can be given the form of a bag model where a broad range
of bag constants $B=75\dots 150$ MeV/fm$^3$ is obtained consistent with 
possible parametrizations of $\pi$ and $\rho$ observables. 

The consequences for deconfinement transition in rapidly rotating neutron 
stars are considered and a new signal from the pulsar timing in binary 
systems with mass accretion is suggested. The model EOS under consideration
meets the new constraints for maximum mass and radius recently derived from
QPO observations. 
Within the present model, quark matter below $T_c\sim 50$ MeV is a 
superconductor of second kind and it is suggested that the magnetic field 
in a neutron star forms an Abrikosov vortex lattice which penetrates into the
quark matter core and thus in accordance with the observation does not decay 
on timescales of $10^4$ years as previously suggested.

\section*{Acknowledgments}
P.C.T. acknowledges  support by the {\sc National Science Foundation} under 
Grant No. INT-9603385 and the hospitality of the University of Rostock where 
part of this work was conducted. 
The work of D.B. has been supported in part by the Deutscher Akademischer 
Austauschdienst (DAAD) and by the Volkswagen Stiftung under grant No. I/71 226.
The authors thank Yu. Kalinovsky, P. Maris, C.D. Roberts and S. Schmidt for 
discussions and criticism.

\end{document}